\documentclass[conference]{IEEEtran}
\usepackage{amsmath,amssymb,amsfonts}
\usepackage{algorithmic}
\usepackage{graphicx}
\usepackage{textcomp}
\usepackage{xcolor}
\usepackage{comment}
\usepackage{tabularx}
\usepackage{graphicx}
\usepackage{url}
\usepackage[export]{adjustbox}
\def\BibTeX{{\rm B\kern-.05em{\sc i\kern-.025em b}\kern-.08em
    T\kern-.1667em\lower.7ex\hbox{E}\kern-.125emX}}
\begin{document}

\title{ChatGPT: A Study on its Utility for Ubiquitous Software Engineering Tasks\\
}

\author{\IEEEauthorblockN{Giriprasad Sridhara}
\IEEEauthorblockA{\textit{Global AI Accelerator (GAIA)} \\
\textit{Ericsson}\\
Bangalore, India \\
giriprasad.sridhara@ericsson.com}
\and
\IEEEauthorblockN{Ranjani H.G.}
\IEEEauthorblockA{\textit{Global AI Accelerator (GAIA)} \\
\textit{Ericsson}\\
Bangalore, India \\
ranjani.h.g@ericsson.com}
\and
\IEEEauthorblockN{Sourav Mazumdar}
\IEEEauthorblockA{\textit{Global AI Accelerator (GAIA)} \\
\textit{Ericsson}\\
Bangalore, India \\
sourav.mazumdar@ericsson.com}
}

\maketitle

\begin{abstract}

ChatGPT (Chat Generative Pre-trained Transformer) is a chatbot launched by OpenAI on November 30, 2022.  OpenAI's GPT-3 family of large language models serve as the foundation for ChatGPT. ChatGPT is fine-tuned with both supervised and reinforcement learning techniques and has received widespread attention for its articulate responses across diverse domains of knowledge.

In this study, we explore how ChatGPT can be used to help with common software engineering tasks. 
Many of the ubiquitous  tasks covering the breadth of software engineering such as ambiguity resolution in software requirements, method name suggestion, test case prioritization, code review, log summarization can potentially be performed using ChatGPT. 

In this study, we explore fifteen common software engineering tasks using ChatGPT. We juxtapose and analyze ChatGPT's answers with the respective state of the art outputs (where available) and/or human  expert ground truth.


Our experiments suggest that for many tasks, ChatGPT does perform credibly and the response from it is detailed and often better than the human expert output or the state of the art output. However, for a few other tasks, ChatGPT in its present form provides incorrect answers and hence is not suited for such tasks.

\end{abstract}


\section{Introduction}
Around 300,000 years ago, the species, homo sapiens (humans) started to emerge in what is the present day continent of Africa~\cite{Hublin2017}. Since then, humans have made spectacular progress in the fields of Science, Technology, Engineering and Mathematics. An indicative but not exhaustive list of these wonderful achievements consists of the discovery of electricity, antibiotics and the DNA; the invention of agriculture, printing press, the number zero and computers;  exploration of space (sending men to the moon), nuclear fission (splitting the atom), the development of the internet and the world wide web.

Certain technology pundits have opined that a Chatbot software released on Nov 30, 2022, was a similar seminal milestone in human achievements.
These pundits have compared the advent of ChatGPT to nuclear fission~\cite{chatgpt-split-atom-article}.

The chat bot known as ChatGPT was released to the public for beta testing by the OpenAI consortium. ChatGPT is a chat bot that can write and debug code, and, automatically repair faulty code. It can compose poems, and attempt medical diagnosis and so on. \footnote{To be fair, ChatGPT does not accept the hyperbolic comparisons. When we queried it as to where it stood in the pantheon of great human achievements, it modestly accepted that the aforementioned achievements ranked significantly higher.}


While we may debate ad infinitum the exact ranking of ChatGPT in the pantheon of human achievements, it is undeniable that it has captured the technology inclined public's imagination like nothing else in recent times. Supposedly it had a million users signing up to use it and millions of conversations with it. Social media platforms are abuzz with users sharing their experience (and screen shots) of their interactions with ChatGPT.
There is anecdotal evidence of incredible responses from ChatGPT such as it passing the US Medical Learning Exam and an exam in the Master of Business Administration offered by the Wharton School in the US. There are also reports, (almost gloating) about ChatGPT s inexplicable failures. For example, apparently, ChatGPT \emph{could not} solve the following elementary algebra problem: `` A bat and a ball together cost 1.10, the cost difference between the bat and ball is 1.0, what are their individual costs?''. 

We are as intrigued as others about the success and failures of ChatGPT. However, as software engineering researchers, our world view is rather limited. We wonder how useful would ChatGPT be for common software engineering tasks.

Towards the above, we catalog a list of fifteen fairly common software engineering tasks spanning the breadth of software engineering. These tasks range across the sub-areas of software development, software quality assurance and software maintenance. We then initiate a conversation with ChatGPT with a specific software engineering task and gauge its response (by comparing with human expert ground truth and/or output from the state of the art tools).

For example, we provide a Java method to ChatGPT and request it to provide a code review of the method. Such code reviews are typical in software development and are often done by experienced developers. We then contrast the provided review with the human expert review. 

We repeat the above experiment with different randomly chosen samples and note the accuracy of ChatGPT. We also repeat the experiment with different tasks such as code summarization, duplicate bug report detection, test case prioritization and so on. We catalog the results noting the number of times ChatGPT does well and where possible, we juxtapose ChatGPT's performance with respect to the human expert output or the state of the art tool output.

The main contributions of this paper are as follows:
\begin{itemize}
    \item A first ever study of ChatGPT and its potential use for fifteen ubiquitous software engineering tasks
    \item A comparison of the ChatGPT output with  human gold set and/or state of the art for the above common software engineering tasks.
\end{itemize}

The remainder of the paper is organized as follows: 
Section~\ref{sec:study} delineates our main study, while, Section~\ref{sec:rel} portrays the related work and we conclude in Section~\ref{sec:conc}.

\section{Study}
\label{sec:study}

In this section, we describe our study. We first describe ChatGPT and the our study setup. We then delineate the experiments for each task. 

\subsection{ChatGPT}

ChatGPT is a large-scale language model developed by OpenAI \cite{chatgpt}, \cite{chatgptblog}. It is trained on a massive dataset of internet text and is able to generate human-like text responses on a wide range of topics. ChatGPT is based on the GPT (Generative Pre-training Transformer) architecture, which has been shown to be highly effective for natural language processing tasks such as language translation, text summarization, and question answering. The model is fine-tuned on specific tasks by fine-tuning the model on a smaller dataset with task-specific examples. 
ChatGPT seems to address a variety of applications including chatbots, language translation and language understanding. It is a powerful tool for natural language generation and understanding, and it has the potential to be used in a wide range of industries and applications.

\subsection{Study Setup} We used the ChatGPT versions released on December 15 2022 and January 9 2023 for our experiments. The study was conducted on our Windows laptop machine of typical configuration.


For each of the tasks, we had a maximum of ten samples. For example, for the task of \emph{Method Name Suggestion}, we interacted ten times with ChatGPT and each time we presented it with a Java method and asked it to suggest a suitable method name. While ten or fewer samples may appear to be very less for a study, it should be noted that this entire study is a completely manual and time consuming task. Further, for many tasks such as \emph{Method Name Suggestion}, we also had to run the state of the art tool with the same input as was fed to ChatGPT and extract the output to juxtapose with the output of ChatGPT. Thus, the number of samples that we have per task is a reasonable sized sample in our opinion.

In each of the following subsections, we describe our experience with using ChatGPT for different software engineering tasks. Each subsection will first describe the task background such as what exactly is anaphora resolution; how we obtained the samples for the experiments; description of the ChatGPT output and comparison with the state of the art. Note that in some cases, there was no state of the art tool but we had human output (gold set), so we compared the ChatGPT output with the human output. Also in some cases, the state of the art tool was not publicly available and neither was a human gold set, so we manually evaluated the ChatGPT output.

For each task, we present numbers on the accuracy of ChatGPT where accuracy is broadly defined as the number of times ChatGPT was correct i.e., its output matched the expected output (from a human or state of the art tool or our manual evaluation). We also present some examples (ChatGPT converation screenshots) of where ChatGPT did well and where it did not do well. We finally summarize the experiment in each subsection and share our opinion on whether ChatGPT is viable for that particular task.



We now start describing each of the experiments that we did in the following subsections.

\subsection{Method Name Suggestion  or Code Summarization}
\textbf{Background:} A method or a function name in any programming language is a succinct summary of what the method does. A well written descriptive method name can help developers in faster  program comprehension as they can chose to read the method body or skip based on what they understand the method does. Thus, it is imperative that a method name describes \emph{precisely} yet \emph{concisely} what a method does.
We wanted to explore if ChatGPT could be used for this common task in software development.

\textbf{Study:} We used the ten publicly available sample Java methods from the companion website
of one of the state of the art tools, \emph{code2seq}~\cite{alon2018codeseq}. We also used the provided web interface of \emph{code2seq} to generate the method name for each of the ten methods. These method names were used to compare with ChatGPT.

\begin{figure}
\centering
\includegraphics[width=\columnwidth, frame]{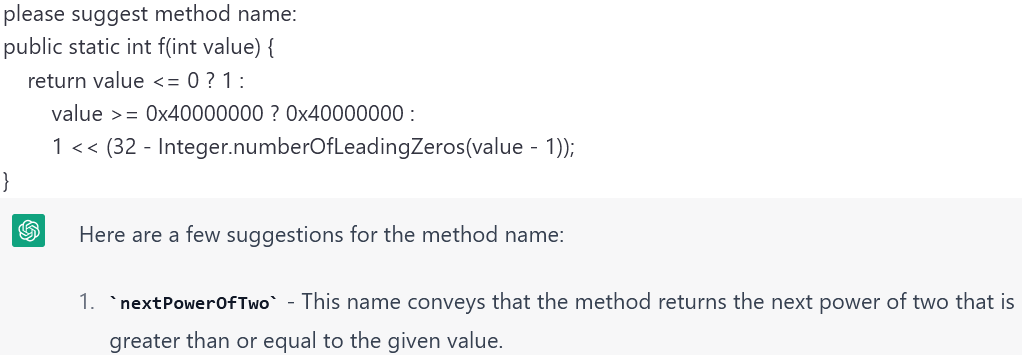}
\caption{Code Summarization : Method Name Suggestion 
}
\label{fig:meth-name-qa-next-pow-2}
\end{figure}

Our chat with ChatGPT was simple: We began by requesting it to suggest a suitable method name for the code, which we copy pasted into our chat. Note that, each method had its name elided and replaced with a meaningless identifier \emph{f}. A sample of our chat is shown in Figure~\ref{fig:meth-name-qa-next-pow-2}.


\textbf{Results:} ChatGPT suggested the correct method name for nine of the ten methods. ChatGPT also suggested better method names for three methods compared to the state of the art. The state of the art generated a better name for one method compared to ChatGPT. These method names are shown in Table~\ref{tab:meth-names-chatgpt-vs-code2seq}.
For the remaining six methods both techniques had method names of similar quality.

\begin{table}[htbp]
\begin{tabular}{ | c | c |}
\hline
ChatGPT & State of Art \\ \hline
saveBitMapToFile & saveBitMap  \\
trackChildRequest & \textbf{addChildRequest} \\
nextPowerOfTwo & getPowerOfTwo \\
\textbf{generateRSAPrime} & generate Prime Number \\
computeStandardDeviation & compute stddev \\
index & index of target \\
index of child & index of item \\
\textbf{BlockTillDone} & check done \\
\textbf{check entity size limit} &report entity size limit \\
waitForJobCompletion & waitForJob \\
\hline
\end{tabular}
\caption{Method Name Suggestions by ChatGPT and State Of Art. Bold text indicates a better name.}
\label{tab:meth-names-chatgpt-vs-code2seq}
\end{table}

\textbf{Observations:} We believe ChatGPT does very well in succinct code summarization i.e., method name suggestion. It generated the correct name nine of ten times. Even for the method where it was not accurate, it only got the action part wrong. It generated \emph{track} instead of \emph{add}, while getting the object of the action, right, i.e., \emph{childRequest}.

\begin{figure}
\centering
\includegraphics[width=0.8\columnwidth, frame]
{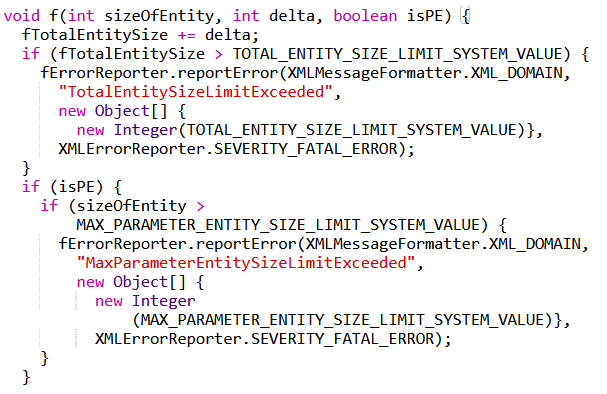}
\caption{Method name suggestion: check (ChatGPT) vs report (state of art)}
\label{fig:method-name-suggestion-check-vs-report}
\end{figure}

Further, as can be seen from the Table~\ref{tab:meth-names-chatgpt-vs-code2seq}, ChatGPT does generate more informative method names for certain methods. For example, consider the Java method shown in Figure~\ref{fig:method-name-suggestion-check-vs-report}.  ChatGPT generates the name \emph{\underline{check} entity size limit}, while the state of the art provides the method name, \emph{\underline{report} entity size limit}. We believe that the method's primary action is to \emph{check} as evinced by the numerous \emph{if} checks in the code. Thus, ChatGPT's method name suggestion is better (more precise) than the state of the art.
Finally, ChatGPT is \emph{noticeably faster} in generating its response than the state of the art.

Therefore, we believe that for the task for method name suggestion or short summary comment generation, ChatGPT performs excellently, generating concise yet precise method names.


\subsection{Log Summarization}
\textbf{Background:} Developers insert log statements at appropriate locations in the source code to capture normal and anomalous behaviours. Logs are created when the software is run. Such logs are extremely useful for developers and system engineers in identifying and fixing different varieties of problems. However, logs can quickly grow in size to thousands of lines. To effectively analyze such logs, some form of summarization is needed. 

\begin{figure}
\centering
\includegraphics[width=\columnwidth, frame]{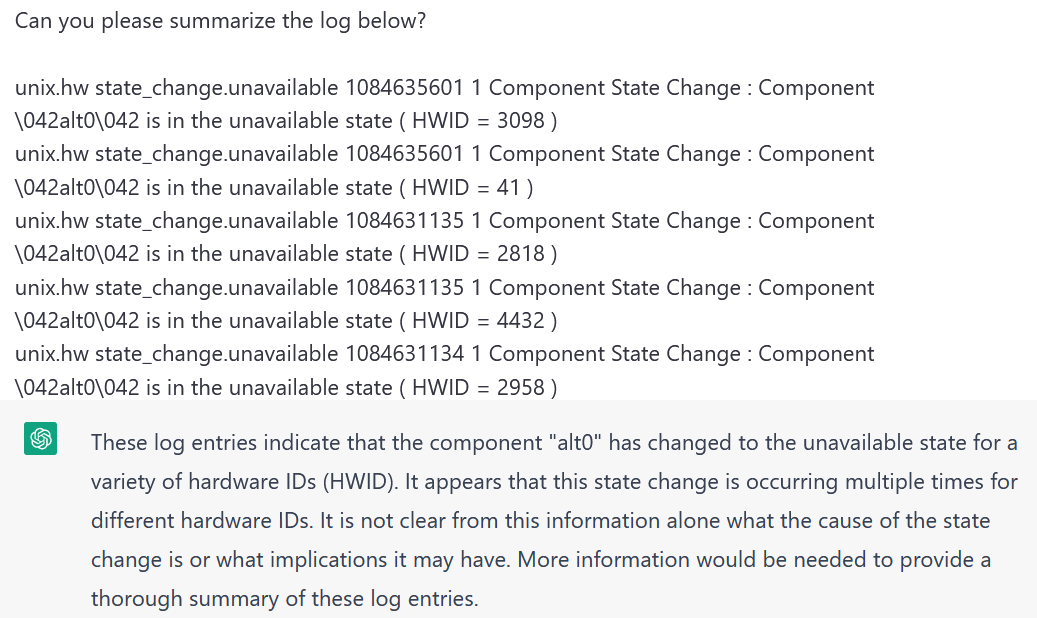}
\caption{Log Summarization}
\label{fig:log-sum-qa-1}
\end{figure}

\textbf{Study:} We wished to understand if ChatGPT could succinctly summarize logs. We gathered ten random logs from the dataset used in a state of the art technique~\cite{log-summary}. These logs have a summary already generated by the technique described in ~\cite{log-summary}. 
Figure~\ref{fig:log-sum-qa-1} shows a sample input log and ChatGPT's summary. Note that the input has been truncated in the screenshot, but it was not truncated in the actual conversation.

\textbf{Results:} ChatGPT generated an appropriate summary for all the ten logs. It also generated a \emph{better} summary than the state of the art all the ten times. 

\textbf{Observation:} ChatGPT does exceedingly well in the task of log summarization, producing summaries that are better than the state of the art. A concrete example is shown in Figure~\ref{fig:log-sum-qa-1}. The ChatGPT summarizes the log as \emph{alt 0 component unavailable} while the state of the art simply summarizes as \emph{state change}. Thus, the ChatGPT summary is more informative without being overly verbose.



\subsection{Commit Message Generation}

\textbf{Background:} Commit messages are succinct summaries of the code changes that were made by a developer for a particular software revision. Commit messages help in faster program comprehension as one can obtain the gist of the changes made without getting into implementation details. However, writing succinct commit messages requires additional manual effort on the part of the developers and due to the multiple commits per day, a developer may not write commit messages. A study has found that almost 14\% of the commits did not have a commit message~\cite{commit-msg-gen-icse-22}.

\begin{figure}
\centering
\includegraphics[width=\columnwidth, frame]{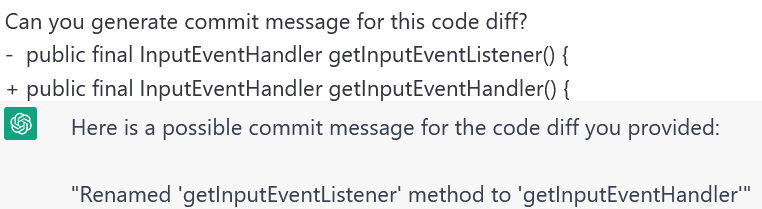}
\caption{Commit Message Generation}
\label{fig:commit-qa-1}
\end{figure}

\textbf{Study:} We wished to ascertain if ChatGPT could generate commit messages. 
Due to the conversational interface of the ChatGPT, we could not provide the entire file versions of the modified files but instead only provided the code diffs as shown in Figure~\ref{fig:commit-qa-1}. We gathered ten diffs from the evaluation dataset of the paper~\cite{commit-msg-gen-icse-22} for use in our experiment.


\textbf{Results:} ChatGPT generated the correct commit message seven of ten times.

\textbf{Observations:}
ChatGPT did generate very good commit messages as shown in the example of Figure~\ref{fig:commit-qa-1}. What was interesting was how seamlessly it understood the ``diff'' language, i.e., lines starting with `-' are removed lines while lines starting with `+' are added lines.
Further it did not generate a naive verbatim message such as \emph{removed getInputEventListener and added getInputEventHandler}. Instead it generated the message like a human would i.e., it understood that the change was a \emph{rename} operation.

We examined the three cases where ChatGPT was not successful. In these cases, ChatGPT was \emph{not} incorrect in its generation but it generated an additional few words. For example, the developer message was: \emph{''Changed the class FormAuthClientBase from private to private abstract''}. ChatGPT generated this part but added \emph{''and inherited it from SimpleHttpClient''}. As the inheritance was already present in the original version, we believe it is not correct to add to the commit message.

Thus, ChatGPT can be used for commit message generation but a perusal of the generated message is needed to remove extraneous content.

\subsection{Duplicate Bug Report Detection}
\textbf{Background:}
Bugs are unfortunately all too common in software. Not only are bugs commonplace, but multiple users can report the same problem. 
Duplicate bug reports are detrimental as they result in precious time being wasted by developers~\cite{duplicate-bug-report-detection-tosem-2022}.


We wanted to determine if ChatGPT could help in duplicate bug report detection. Typical duplicate bug detection involves comparing a given bug report against a set of existing bug reports and checking if a duplicate exists. However, due to the ChatGPT interface, we modified the problem by providing two bug reports to ChatGPT and asking it if the bug reports were duplicates of each other.

\begin{figure}
\centering
\includegraphics[width=\columnwidth, frame]{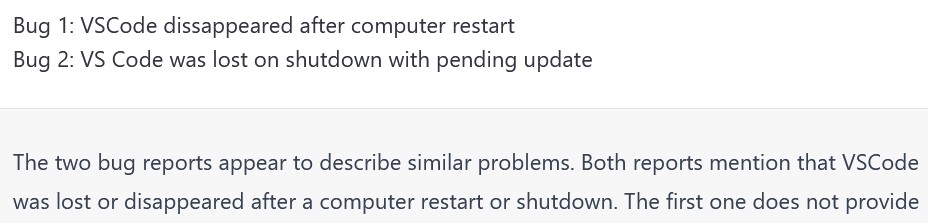}
\caption{Duplicate Bug Report Detection}
\label{fig:duplicate-bug-report-correct-q-a-1}
\end{figure}

\textbf{Study:} We gathered ten bug reports marked as already having a prior duplicate bug from the publicly available bug report database of Microsoft VSCode (a popular source code editor)~\cite{duplicate-bug-report-dataset-ms}. 
We provided pairs of duplicate bug reports to ChatGPT as input and asked it to decide whether the bug reports were duplicates.
A sample of our interaction with ChatGPT is shown in Figure~\ref{fig:duplicate-bug-report-correct-q-a-1}. 
(Note that due to space constraints, we have shown only the titles of the bugs in the figure. Although, in our actual converation, we provided both the title and the details.)

\textbf{Results:} Out of the ten pairs of duplicate bug reports, ChatGPT correctly identified six pairs as being duplicates of each other.

\textbf{Observations:}  In Figure ~\ref{fig:duplicate-bug-report-correct-q-a-1}, ChatGPT was able to identify the duplicates although the bug reports used different terms to describe the same problem. 
However, in the four cases where ChatGPT failed, it perhaps was surprisingly confounded by the use of different terms. For example, in one pair of duplicate bug reports, the first report was \emph{ Unable to copy and paste in output terminal bash}, while, the second report was \emph{Terminal has no response to typing on Codespace}. 

Thus, the results are largely underwhelming. ChatGPT is perhaps not fully suited to the task of duplicate bug report detection in its present release.


\subsection{Merge Conflict Resolution}

\textbf{Background:} In modern software development, multiple developers can often work on the same unit of code such as a class or a file. To avoid conflicts, developers create their own branch in a version control repository such as \emph{git} and make their changes. Periodically, changes from the different branches have to be synchronized or merged into the main or master branch. Unfortunately, multiple developers may make conflicting changes around the same lines of code. It is estimated that around 46\% of merges result in a conflict. Resolving conflicts manually is a tedious and error prone activity~\cite{merge-conflict-resolve-fse-22}.
We explore if ChatGPT can automatically resolve merge conflicts.

\textbf{Study:} We took ten random samples from the evaluation dataset provided by ~\cite{merge-conflict-resolve-fse-22}. With each sample, we asked ChatGPT to resolve the merge conflict by providing the following:
\begin{enumerate}
    \item A base version of the file which has code common to both changes that are being merged. 
    \item Content of the file having one set of changes to be merged (Version 1)
    \item Content of the file having the \emph{other} set of changes to be merged (Version 2)
\end{enumerate}

\begin{figure}
\centering
\includegraphics[width=\columnwidth, frame]{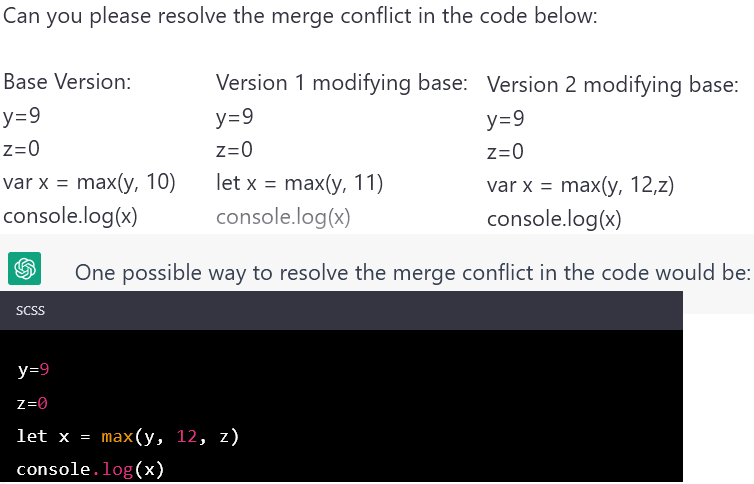}
\caption{Merge Conflict Resolution}
\label{fig:merge-qa-1}
\end{figure}

A sample question and ChatGPT's answer are shown 
in Figure ~\ref{fig:merge-qa-1}.
To compute ChatGPT's accuracy, we juxtaposed ChatGPT's output with the actual final resolved version (the original developer's).


\textbf{Results:} ChatGPT was successful 6 of 10 times.

\textbf{Observation:} When it was successful as in Figure ~\ref{fig:merge-qa-1}, ChatGPT was impressive, deftly merging conflicting tokens. It also displayed human intelligence in some other situations. For example, the copy right year was 2019 in one version and 2020 in another. ChatGPT used the latest 2020 in its resolved merge operation.

In the four cases where it failed, one sample was truly difficult to merge as the two versions had modified a method's return type. In the other cases, ChatGPT simply did a union of the changes in the two versions which did not match with the developer's merge operation.

To conclude, ChatGPT perhaps needs more training to work better on resolving merge conflicts.

\subsection{Anaphora Resolution}
\textbf{Background:} 
Natural Language (NL) is commonly used to specify the software requirements, constraints under which it has to operate, performance expectations and so on. NL is unfortunately prone to issues of ambiguity. Ambiguity occurs when a word has multiple interpretations. An especially pervasive form of ambiguity 
is the one that occurs due to anaphoras.

\emph{Anaphora} is essentially a reference to entities mentioned earlier in the text. Typical examples of anaphors are pronouns such as \emph{it, he, she}. Anaphora resolution is the task of detecting the entities to which an anaphor refers to.


\textbf{Study:}
We gathered a random set of ten requirement specifications worded in English and needed anaphora resolution from the experimental dataset provided by one of the latest works on anaphora ambiguity~\cite{anaphora-ambiguity-icse-22}. These requirement specifications had anaphora ambiguity and the resolution annotated by two human experts.

\begin{figure}
\centering
\includegraphics[width=\columnwidth, frame]{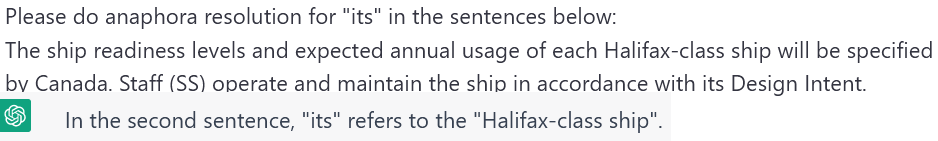}
\caption{Anaphora Resolution}
\label{fig:anaphora-qa-1}
\end{figure}

We asked ChatGPT to perform anaphora resolution for each of the ten requirement specifications and compared its output with the gold set output of the human experts. A sample of our question and ChatGPT's answer is shown in Figure~\ref{fig:anaphora-qa-1}.

\textbf{Results:} ChatGPT correctly identified the \emph{antecedent} to which the anaphora referred to in \emph{all} the ten instances.



\textbf{Observation:} For the task of anaphora ambiguity resolution, ChatGPT is extremely viable.

\subsection{Code Review}

\textbf{Background:} Code review by experienced developers is an inalienable part of software quality assurance (along with static analysis and testing).
Code review can find errors that are not found by testing or static analysis. For example, potential vulnerabilities in the libraries used. Code review requires experienced developers who unfortunately may not have the time to review code due to their everyday activities. 

\begin{figure}
\centering
\includegraphics[width=\columnwidth, frame]{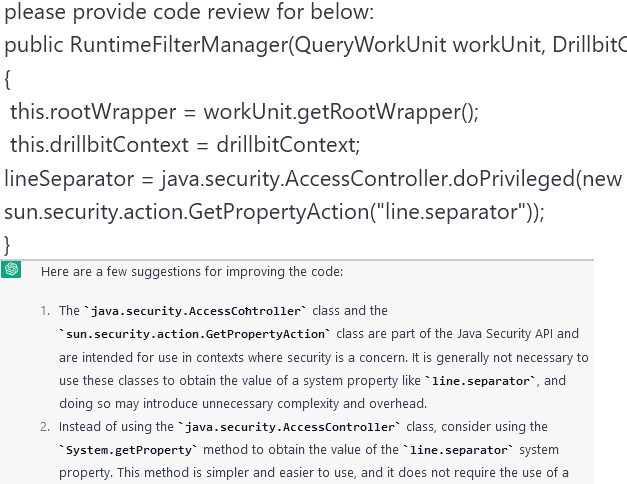}
\caption{Code Review}
\label{fig:code-review-qa-1}
\end{figure}

\textbf{Study:} We collected ten random methods from a recent paper on automating code review~\cite{code-review-icse-22}.
We provided each method to ChatGPT and requested it to generate a code review. We contrasted the generated review with the original developers' review present in the dataset. Figure \ref{fig:code-review-qa-1} shows a sample code review generated by ChatGPT.



\textbf{Results:} There was agreement between ChatGPT's generated code review and the original developer's code review on only four of the ten cases.

\textbf{Observation:} ChatGPT started impressively for code review generation as depicted in Figure~\ref{fig:code-review-qa-1}. Particularly prominent was how it flagged the misuse of the \emph{Java Security API AccessController} class and the recommendation to use the more benign \emph{System Property} to get the new line property. This review matched exactly with the developer's review.

Sadly, ChatGPT did not do so well on the remaining samples.
We considered the possibility that the original developer had the benefit of domain knowledge about the project which enabled him/her to provide an insightful review comment which could not be replicated by ChatGPT as it was bereft of the domain knowledge.

However, when we examined the cases where ChatGPT did not match the developer's review, the discrepancy could not simply be ascribed to lack of domain knowledge about the code.
For example, in one method, a message about a severe error was logged using the lowest logging level. The original developer review asked the level to be modified to the highest logging level viz., \emph{error} level. ChatGPT did not identify this. 

Thus, we conclude that ChatGPT in its current form cannot be used for code review generation.

\subsection{Type Inference Suggestion}

\textbf{Background:} 
Dynamic typing that is allowed in languages like Python enables faster prototyping and hence has found resonance among developers; especially those developing machine learning applications. However, dynamic typing can lead to issues in security, performance and program comprehension. Hence, some kind of type hints about a variable are desired strongly by the developers
~\cite{python-type-inference-icse-22}.

\begin{figure}
\centering
\includegraphics[width=\columnwidth, frame]{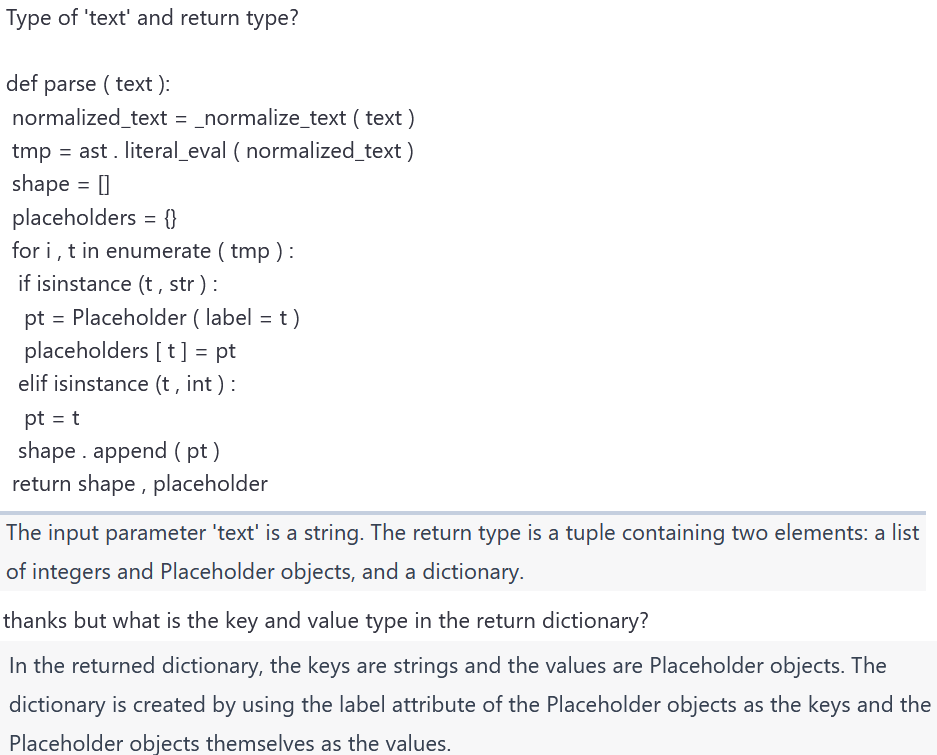}
\caption{Type Inference}
\label{fig:type-inf-qa-1}
\end{figure}

\textbf{Study:} We used ten random Python functions from the dataset of ~\cite{python-type-inference-icse-22} for the study.
We asked ChatGPT about the type of certain variables in the given function. A sample is shown in Figure~\ref{fig:type-inf-qa-1}.
The dataset had the ground truth i.e., the type hints for the variables.



\textbf{Results:} ChatGPT succeeded seven out of ten times.

\textbf{Observations:}  Figure~\ref{fig:type-inf-qa-1} demonstrates an impressive type inference done by ChatGPT. When we initially saw the response, we thought that ChatGPT was incorrect about the return list containing both integers and placeholder objects. After reading the function multiple times, we discovered that ChatGPT was correct. 

We also examined the cases where ChatGPT was incorrect. In one case, ChatGPT inferred the type of a variable as an array in the numpy library while in actuality, it was an array in another library. In another case, ChatGPT was not able to correctly identify the return type as it was a function call and it did not have access to the function body.




To summarize, ChatGPT while not being perfect can be a good adjunct tool for type inference.

\subsection{Code Generation from Natural Language for Data Frame Analysis}
\begin{figure}
\centering
\includegraphics[width=\columnwidth, frame]{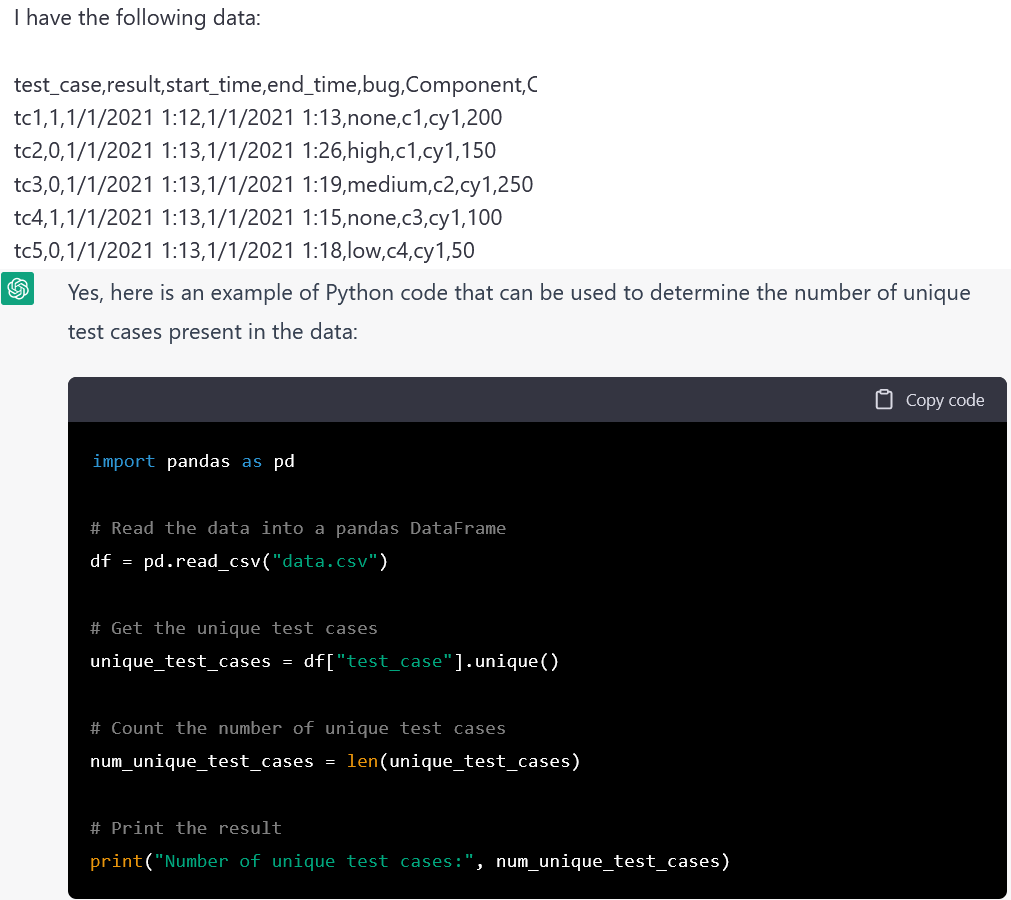}
\caption{Code Generated by ChatGPT as a response to "Generate code to determine number of unique test cases."}
\label{fig:df-nl-codegen-qa-1}
\end{figure}

\textbf{Background:}
One common work activity that many scientists, especially, data scientists do is analyze data. Typically, data is in the form of a table as shown for example in the top part of Figure~\ref{fig:df-nl-codegen-qa-1}. 

Data Scientists analyze the data by writing code in Python (often with the \emph{numpy} and \emph{pandas} library). We wanted to explore how ChatGPT would perform in generating data analysis code given natural language requirements.

\textbf{Study:} We provided a tabular data set to ChatGPT and asked it to generate Python code for the questions shown below:
\begin{enumerate}
    \item  How many unique test cases are present?
    \item  Which test case has maximum failures?
    \item  What is the average execution time of each test case?
    \item  Which test case leads to the maximum high priority bugs?
    \item  Write code to check if there is a correlation between execution time and whether the test passes or fails.
    \item  Generate code to determine correlation between execution time and pass or fail using Spearman or Pearson correlation.
    \item \textbf{For the given data, can you generate a machine learning classifier to predict whether a test run passes or fails?}
    \item Can you use test case name and execution time as feature?
    \item \textbf{Can you generate code for test case prioritization based on past results?}
    \item Write code to get bug count per component per priority.
\end{enumerate}

Note that the data frame and the questions are not the standard tutorial dataframe and questions that one can find on the Internet. Thus, the answers of ChatGPT i.e., the code it generates is perhaps not something that is copied directly from the Internet.

Once ChatGPT generated its response which included pandas code, we copy pasted the code into a Jupyter notebook and validated the code by executing it. 

\textbf{Results:} ChatGPT generated the correct functioning code in eight of the ten cases.
The two questions marked in bold above represent the failure cases.

\textbf{Observations:} ChatGPT correctly recognizes the prevailing paradigm of using the \emph{pandas} library for such data analysis in Python. It generates appropriate code with well named and meaningful identifiers. The produced code is also well commented as shown in Figure~\ref{fig:df-nl-codegen-qa-1}.

Further, even more impressive is the fact that to compute average execution time, it understands that the actual execution time per run must be found and computes this as the difference of the columns \emph{end time} and \emph{start time}. 
It also understood that in this particular data frame, 0 represented failure of a test run (typically, the paradigm in test case executions is to use 0 for success and 1 for failure).

However, for two of the questions, the code generated by ChatGPT was not satisfactory. These are the highlighted questions above.

For the first of these questions, we asked it to generate code to learn a machine learning classifier to predict which test run would pass or fail. In our data set, there were two features that would help find the \emph{segregating hyperplane} between pass and fail, viz., \emph{test case name} and \emph{execution time}. However, ChatGPT \emph{did not} select these as features initially and only after prompting by us via a subsequent question, did it chose them as features.

ChatGPT also did not generate correct code for the test prioritization question. Here it assumed that a result value of 1 stood for failure and thus produced code which would result in a test prioritization order that was the \emph{opposite} of the correct order. This was surprising because to answer a previous question about faults, it had correctly generated code to count the zeros to denote total faults.


To summarize, ChatGPT does very well in understanding the natural language questions on a dataframe and generates suitable code. It perhaps suffers a bit in feature engineering i.e., in deciding which features are important to learn a good classifier that can distinguish between the positive and negative samples.

\subsection{Defect or Vulnerability Detection}


\textbf{Background:} 
Given lines of source code, the task is to identify whether it is an insecure code that may attack software systems, such as resource leaks, use-after-free vulnerabilities and Denial of Service attacks. This is essentially a binary classification problem.

\begin{figure}
\centering
\includegraphics[width=\columnwidth, frame]{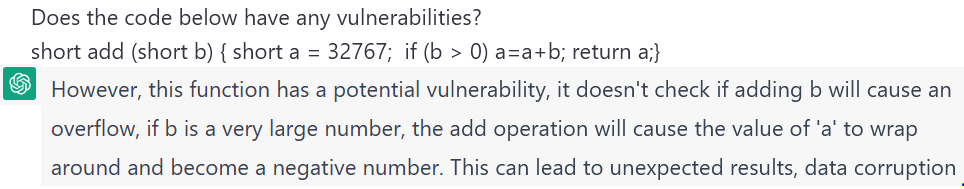}
\caption{Vulnerability Detection}
\label{fig:vuln-qa-1}
\end{figure}

\textbf{Study:} We took ten random methods that were marked as having vulnerabilities from the dataset in~\cite{vuln-devign}. We provided each method to ChatGPT and asked it to identify if the method had a vulnerability as shown in Figure~\ref{fig:vuln-qa-1}.


\textbf{Results:} ChatGPT correctly identified the vulnerabilities in only four of the ten cases. 

\textbf{Observations:} It should be noted that vulnerability detection is a relatively difficult task requiring extensive program analysis (control and data flow) which perhaps ChatGPT is not trained for. 

Perhaps the relatively low level nature of the code snippets which are C programs written with  pointer access and indirection might have been the reason as to why ChatGPT was not able to identify the vulnerabilities.

To conclude, we perhaps cannot use ChatGPT \emph{as is} for vulnerability detection. 

\subsection{Code Clone or Duplicate Code Detection}
\textbf{Background:} Code clones are code snippets that accomplish a similar functionality.
Code clones may have similar or different syntax 
(for example, one snippet may loop using \emph{for}, while the other snippet may loop using \emph{while}). Code clones can increase maintenance costs due to duplication of bugs. Hence there are several approaches to automated code clone detection~\cite{big-clone-dataset}.

\begin{figure}
\centering
\includegraphics[width=\columnwidth, frame]{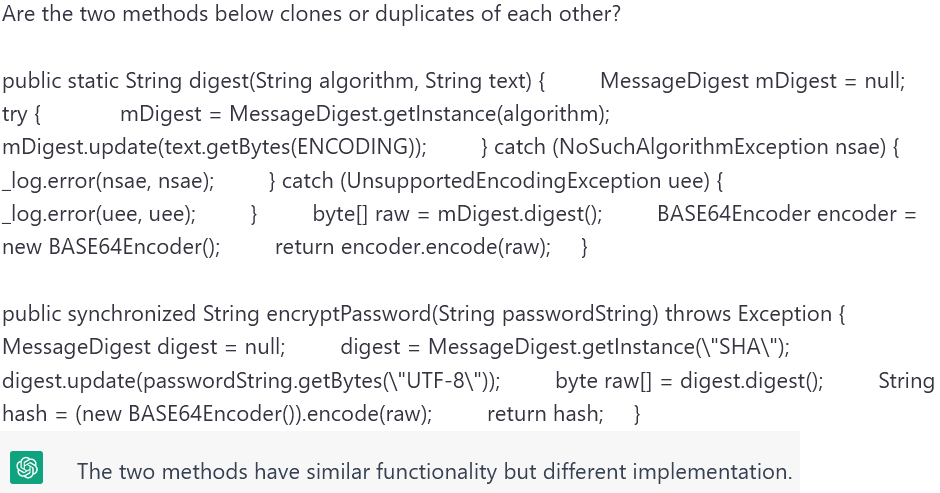}
\caption{Code Clone Identification}
\label{fig:clone-qa-1}
\end{figure}

\textbf{Study:} We gathered ten pairs of Java functions marked as clones of each other from the BigCloneBench dataset~\cite{big-clone-dataset}. We provided the methods as input to ChatGPT and requested it to identify if the methods were clones of each other. A sample is shown in Figure~\ref{fig:clone-qa-1}.

\textbf{Results:} ChatGPT was able to correctly identify six of the ten code clones. Notice how it was able to identify the clones in Figure~\ref{fig:clone-qa-1} despite the methods having different names and parameters and differing implementation.

\textbf{Observation:} We examined the four cases where ChatGPT erroneously said that the method pairs were \emph{not} similar. We found that in three cases, ChatGPT was actually correct and the ground truth set was incorrect. In the remaining case, it was a matter of semantics. The input methods were both named `copyFile' but ChatGPT argued that there were \emph{not} duplicates because the second method was actually 'copyFile\textbf{s}' i.e., it was copying multiple files if the input source file was a directory (via a recursive call).

Thus, we believe ChatGPT does well in the code clone detection and is able to make fine grained distinctions and cogent arguments about its decisions.


\subsection{Test Oracle Generation}

\textbf{Background:}
Software testing can be done on the entire software (functional testing) or can be done at smaller granularity such as testing of a unit, where a unit is a class (file) or a method (function). 

There is considerable interest and research on automating unit tests, especially, automatically generating a unit test, given a method under test (\emph{focal method}) with its signature and body. 

Unit testing typically involves invoking the \emph{focal method} with suitable parameters and then checking the return value. The return value is checked using \emph{assert} statements.

One challenging aspect of unit test generation is to automatically generate the correct assertions (test oracle), as the developers may not always be completely familiar with the method under test.

\begin{figure}
\centering
\includegraphics[width=\columnwidth, frame]{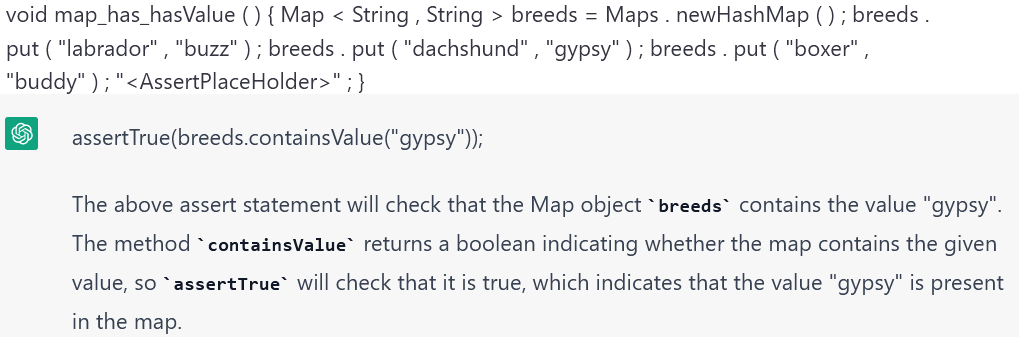}
\caption{Test Oracle Generation}
\label{fig:assert-qa-1}
\end{figure}

\textbf{Study:} We wanted to inspect if ChatGPT could help in the task of test oracle generation. Towards this goal, we collected ten random samples from the state of the art paper described in ~\cite{test-oracle-assert-gen-icse-22}. These samples had the method under test and the test code with a place holder for the assert statement to be generated as shown in Figure ~\ref{fig:assert-qa-1}. We requested ChatGPT to generate a suitable assert statement. We then compared ChatGPT's assert statement with the ground truth which was the original developer assertion.



\textbf{Results:}
ChatGPT's assert statement matched the ground truth six of ten times.

\textbf{Observations:} 
ChatGPT not only generated the assert statement but also provided a cogent explanation of the assertion as shown in Figure~\ref{fig:assert-qa-1}. 

We examined the four cases where ChatGPT failed and found that in one instance, the test method was testing a vector dot product operation while the focal method was the Vector class constructor. ChatGPT identified this anomaly correctly. Considering the other failure cases, ChatGPT generated the opposite assertion (\emph{assertTrue (m.matches())} as opposed to \emph{assertFalse(m.matches())}. In another case, it erroneously used a variable declared within the focal method in the assert statement of the test method.

Thus, we believe ChatGPT's performance was average in the task of oracle generation.
However, with some manual supervision, it can be utilized to generate assert statements.

\subsection{Code Refactoring}
\label{sec:refactor}

\textbf{Background:}
Code refactoring is a software maintenance activity in which  developers modify the code to make it more readable and hence increase maintainability~\cite{refactoring}. Refactoring can be done at a class level, method level and so on. One particular method level refactoring technique is ExtractMethod refactoring in which portions of code from an existing method or function is extracted into another method, while replacing the original code with a call to the extracted method.

\textbf{Study:} We gathered different ExtractMethod refactorings performed by developers from the dataset in ~\cite{refactoring} and asked ChatGPT to perform ExtractMethod refactoring. We contrasted the modified code with the refactoring done by the original developers.



\textbf{Results:} Refactorings suggested by ChatGPT did not agree with the original extract method refactorings for any of the samples. 

\textbf{Observations:}
We manually examined the ExtractMethod refactorings produced by ChatGPT.
While they did not match with the developer's refactorings, we found them to \emph{syntactically correct and semantically matching} the original method's computational intent. 

We also found the ChatGPT's refactored method more readable compared to the original unrefactored method. ChatGPT also provided good identifier names to the extracated methods, its parameters and the return value.

ExtractMethod refactoring is a \emph{subjective} activity, the decision on which statements should be extracted into another method is inherently subjective in nature. Thus, the fact that the method extractions performed by ChatGPT does not match with the developer's refactorings should not preclude its usage for refactoring.

\subsection{Natural Language Code Search}
\textbf{Background:} Developers often have to read and understand code written by other developers to fix bugs and/or add new features. In large code bases with thousands of files and millions of lines of code, developers resort to searching the code base to understand how and where a feature is implemented. While such an activity is a classical information retrieval search and rank problem, we can also formulate the problem as a binary classification problem: Given a natural language query such as "create directory using absolute path", determine if a snippet of code "answers" the query i.e., the code performs the task(s) mentioned in the query. 

\begin{figure}
\centering
\includegraphics[width=\columnwidth, frame]{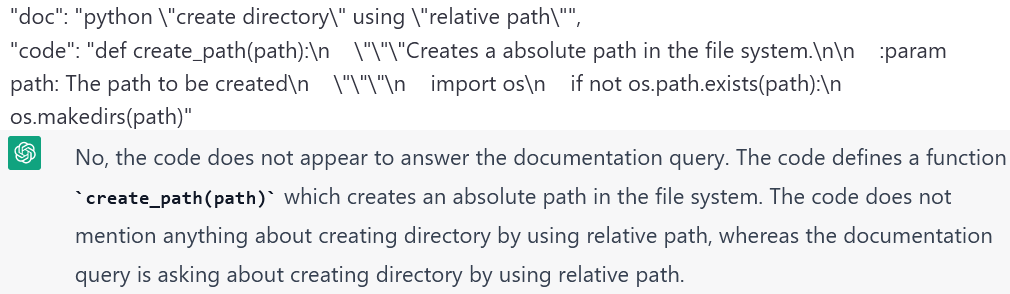}
\caption{Code Search}
\label{fig:code-search-qa-1}
\end{figure}

\textbf{Study:} We took ten random pairs of natural language queries and an associated code snippet from the CoSQA dataset (Code Search and Question Answering)~\cite{huang-etal-2021-cosqa}. Each pair had a ground truth label of 1 or 0, depending on whether the code answered the query or not, respectively. We then provided the natural language query and the code snippet to ChatGPT and asked it if the code answered the query, as shown in Figure~\ref{fig:code-search-qa-1}. We contrasted the ChatGPT output with the ground truth in the CoSQA dataset.


\textbf{Results:} The output of ChatGPT matched  the ground truth \emph{seven} of \emph{ten} times. 

\textbf{Observation:} 
ChatGPT was able to 
make fine-grained and subtle distinctions of meaning as exemplified in Figure~\ref{fig:code-search-qa-1} (between "relative" and "absolute" paths). Upon inspection of the three failed cases, we found that in one case the ground truth had an error. The documentation asked for code to check a ctypes array whereas the code checked for a C++ array, which ChatGPT explained was different. In another case, we believe the natural language query was poor and not informative enough ( "ask if a method can see a variable python"). \emph{Therefore, for code search ChatGPT could be used as an adjunct tool.}.

\subsection{Test Case Prioritization}
\textbf{Background:} In modern software development, we typically have thousands of test cases. Due to the frequent updates to the code base (often multiple times per hour), it is not feasible to run all the test cases for every update. Thus, a ranking scheme to prioritize test cases that are more likely to fail is desired and several automated approaches have been proposed for test case prioritization ~\cite{tcp-rl-2020}. 

Broadly, we explored if ChatGPT can help with test prioritization using three different approaches which are each described below:

\begin{enumerate}
    \item Prioritization based on past faults found
    \item Prioritization based on code changes
    \item Prioritization based on operation order
\end{enumerate}

\subsubsection{Prioritization based on past faults found}
\begin{figure}
\centering
\includegraphics[width=\columnwidth, frame]{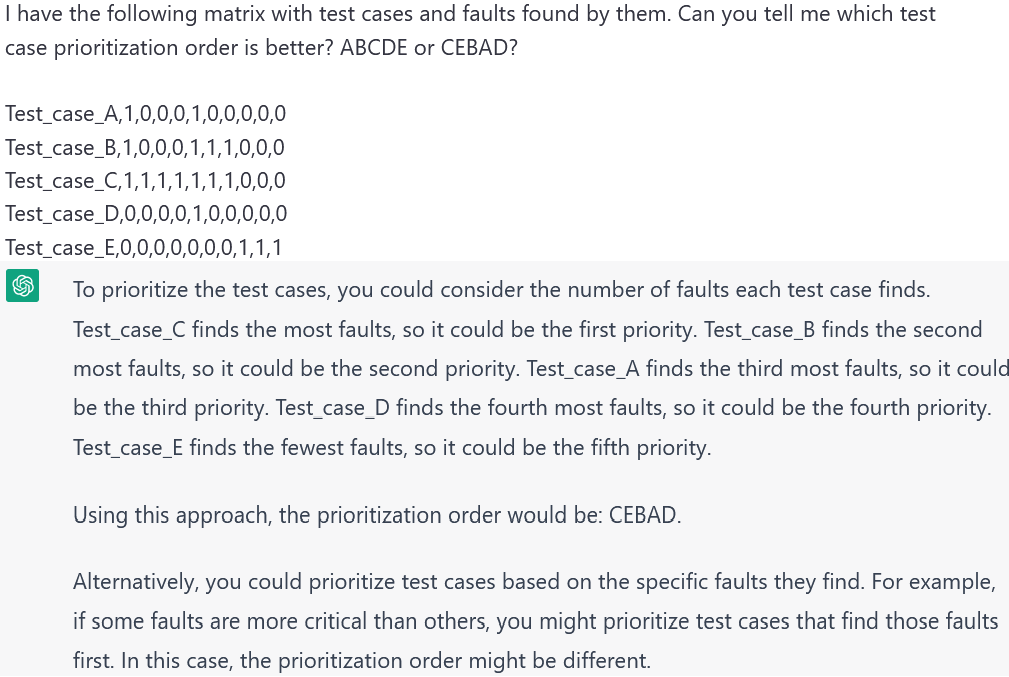}
\caption{Test Case Prioritization}
\label{fig:tcp-qa-1}
\end{figure}

One approach to prioritization is to use the previous or historical fault(s) found by a test case. We provided the data as shown in the top half of Figure~\ref{fig:tcp-qa-1}. The data shows five test cases viz., A to E and ten faults. Note that a fault can be detected by different test cases. For example, the first fault is found by test cases, A, B and C, while the last (tenth) fault is found only by test case E. We asked ChatGPT as to which prioritization order of the test cases was better. 
For this data, executing test case C and then test case E, ensures that all the faults are detected and thus C and E should be ranked at the top.

As can be seen from the lower half of the Figure~\ref{fig:tcp-qa-1}, ChatGPT provided the correct order \emph{CEBAD}, although the reasoning it provided was not consistent with its answer. The reasoning of ChatGPT is based on the \emph{naive} way of prioritization, which favors test cases which find most faults. However, such an ordering is often suboptimal, as in this case.

\subsubsection{Prioritization based on code changes}
\label{sec:tcp-ir}
Another approach to test prioritization is to rank based on what changed in the code. Here, we take the code diffs and use it as a query against each test case and rank based on a similarity measure such as cosine similarity. This is an Information Retrieval based approach~\cite{ir-tcp-issta-2020}.

We provided sample code diffs from the above paper's dataset and also each test case as input to ChatGPT and requested it to prioritize based on textual similarity. However, ChatGPT could \emph{not} perform this task, which is surprising since it did perform code clone detection, duplicate bug report identification which were also somewhat similar tasks.

\subsubsection{Prioritization based on operation order}
Modern software architecture especially the microservices architecture largely relies on REST APIs~\cite{rest-api-testing}. These APIs perform operations such as \emph{add item to cart}, \emph{checkout} and so on. The specifications of these APIs such as the parameters they take, the constraints on them, the return values and so on are stored in a JSON file conforming to a specification called Swagger specification.

Within the REST APIs for an application, there is typically an ordering among the APIs.
For example, the API, \emph{addToCart} should precede \emph{checkout}. A partial order among the APIs can be obtained by analyzing the Swagger JSON.

We provided ChatGPT with a Swagger JSON for a sample application (Petstore) and asked it to obtain the partial order among the APIs. While it did understand the Swagger specification, unfortunately, it was not able to prioritize amongst the APIs.

\textbf{Observation:} 
ChatGPT can be used with prioritization based on past faults found. However, for other kinds of prioritization it perhaps is not viable in its current state.

\subsection{Other Tasks}
We experimented with a few more tasks which we describe below:

\subsubsection{Code Deobfuscation}
JavaScript code is typically downloaded and to reduce download size, a name like \emph{nextPowerOfTwo} can be \emph{minified or uglified} to a name like \emph{f}. However, we may need to recover the original name for comprehension, security review and so on. 
We wanted to explore if ChatGPT can help in this task.

We requested ChatGPT to deobfuscate the illustrative example JavaScript method shown in ~\cite{deobfuscation}. However, ChatGPT was not able to recover the original names. Thus, for this task of deobfuscation, ChatGPT may perhaps not be useful.

\subsubsection{Efficient Algorithm Generation:} Often developers have to write efficient code that takes less time and space. For example, device an algorithm that runs in O(n) time.
We wanted to explore if ChatGPT could help in this task. We gathered a few sample problems with such time complexity constraints from the popular competitive programming website \emph{LeetCode}~\cite{leetcode} and asked ChatGPT to solve the problem. 
ChatGPT was able to generate code which matched the ground truth solutions. In future work, we will evaluate this task with the standard ten samples.

\subsubsection{Automated Program Repair:}
Here the task is to automatically fix bugs in the code, thus reducing the cost of bug-fixes. We took a few Java methods with bugs from the CodeXGLUE dataset~\cite{codexglue} and asked ChatGPT to automatically fix the bugs. However, ChatGPT was not able to perform this task because the input method had \emph{normalized} all the identifier names (for example, \emph{ private TYPE 1 getType ( TYPE 2 VAR 1 ) } ). In future work, we will evaluate this task using the original methods with bugs.


\subsection{Overall Summary:}
Table~\ref{tab:chat-gpt-success-rate} summarizes the overall performance of ChatGPT. 


\subsection{Threats to Validity}

The number of samples used 
in the study might be considered small. However, as explained before in Section~\ref{sec:study}, we perforce had to limit the number of conversations we had with ChatGPT as the study is a completely manual and time consuming activity.

It is possible that some might consider that the tasks themselves are not representative of the kind of activities done in software engineering. To address this concern, we chose the tasks based on multiple different factors such as domain knowledge of the industry; examination of the activities done in several popular open source projects; the research trends shown in the academia and industry.

\begin{table}[htbp]
\begin{tabular}{ | c | c | c |}
\hline
Task & Success & Failure \\ \hline
Method Name Suggestion  & 9 & 1 \\
Log Summarization  & 10   & 0 \\
Anaphora resolution  & 10   & 0 \\
Python Type Inference   & 7 & 3 \\ 
Commit Message Generation   & 7 & 3 \\
Code Review   & 4 & 6 \\ 
Duplicate Bug Report Detection   & 6 & 4 \\
Natural Language Code Search   & 7 & 3 \\ 
Vulnerability Detection   & 4 & 6 \\
Code Clone Detection   & 6 & 4 \\
Test Oracle Generation    & 6 & 4 \\
Code Generation from NL  & 8 & 2 \\
Merge Conflict Resolution & 6 & 4 \\
Code Refactoring*   & 10  & 0 \\ 
Test Prioritization ** & - & - \\

\hline
\end{tabular}
\caption{ChatGPT: Performance across Software Tasks. * denotes \emph{manually verified for syntactic and semantic correctness}.
** denotes that ChatGPT was not able to perform the task at all across multiple samples.}
\label{tab:chat-gpt-success-rate}
\end{table}

\section{Related Work}
\label{sec:rel}


CodeBERT~\cite{codebert} was the earliest bimodal pre-trained model for NL and programming language (PL). It was pre-trained on 6 programming languages. The objective functions included Masked Language Model (MLM) on bimodal data of NL-PL pairs and Replaced Token Detection (RTD) \cite{clark2020electra} on both unimodal and bimodal data and evaluated on downstream tasks including natural language code search and code-to-documentation generation. 

CodeXGLUE \cite{codexglue, codexglueLeaderBoard} is an evaluation benchmark released by Microsoft for General Language Understanding Evaluation benchmark for CODE. It includes 14 datasets for 10 diversified PL tasks covering code-code (clone detection, defect detection, cloze test, code completion, code refinement, and code-to-code translation), text-code (NL code search, text-to-code generation), code-text (code summarization) and text-text (documentation translation) scenarios. 



Codex is OpenAI's natural language to code AI model \cite{codexarxiv2021,codexReleaseNote, codexBlog}. It is trained for the task of generating standalone Python functions from docstrings, and evaluate the functional correctness of code samples automatically through unit tests. The work also introduces HumanEval, an evaluation set to assess programming language comprehension, reasoning, algorithms, and simple mathematics. Evaluation of several Codex models (parameters ranging from 12M to 12B), shows that a compact Codex-300M version outperforms GPT-J-6B. Limitations of large training data, challenges with long specifications, syntactically incorrect code apart from legal, environmental risks associated have been discussed. The Codex model is mostly associated with Copilot tool of GitHub \cite{codexReleaseNote}. The differentiating factor relates to integration of Copilot in a development IDE \cite{copilotReleaseNote}.  However, it is well acknowledged that an independent evaluation of either Codex or Copilot will be challenging owing to lack of visibility of the training data \cite{frank2022llmforcode}. From a human usability perspective, a recent study \cite{copilotstudy}, conducted on 24 participants, reports that although users did use Copilot as a starting point in daily programming tasks, challenges in understanding, editing, and debugging these code snippets hindered their task-solving effectiveness.

In early 2022, OpenAI released InstructGPT \cite{instructgpt}. This LLM was trained on the task to act in accordance with the user’s intention. For this, reinforcement learning with human feedback (RLHF) was used with GPT-3 model to follow a broad class of written instructions. It was found that users preferred outputs from InstructGPT over GPT-3. 

EleutherAI released The Pile, an 825 GB English text corpus for building large-scale language models (LLM) \cite{thepile} in 2020. As an alternative to OpenAI's GPT-3, open-source LLMs such as GPT-Neo and GPT-J \cite{gpt-j, gpt-neo} were released. In 2022, GPT-NeoX, a 20B parameter LLM was released. All these models were trained on The Pile and are available freely to the public.

The work of \cite{frank2022llmforcode} evaluates six large language models of code including Codex, CodeGen and the open-source variants (such as GPT-J, GPT-Neo and GPT-NeoX trained on code) alongside their proposed open-source large language model of code (trained on 12 programming languages). The chosen tasks are code completion and code synthesis from natural language description. Their evaluations show that Codex variant with 300M parameters shows promise on the HumanEval dataset and that open source LLM for code still have a lot of room for improvement. 

We think, the ChatGPT in its current form could be a consolidation of OpenAI's various code-to-text and text-to-text initiatives along with their recent advances in learning to act according to user's intent. 
To the best of our knowledge, ours is the first research work on ChatGPT to do a rigorous study across different software tasks. 

\section{Conclusion}
\label{sec:conc}

In this paper, we asked the question "What is the utility of ChatGPT towards common software engineering tasks such as code summarization, test oracle generation and so on?"

To answer the above question, we chose fifteen ubiquitous tasks in software development and conducted a study to check the feasibility of ChatGPT towards helping with these tasks. We gathered ten random samples for each task such as code clone detection and asked ChatGPT via its chatbot interface to perform the desired task (for example, answer if the provided code snippets were duplicates or clones of each other). We then compared the answer of ChatGPT with the human expert output and/or state of the art tool output. We computed the accuracy of ChatGPT for each task.

ChatGPT does very well on the tasks of log summarization, anaphora resolution, code summarization (method name generation) and code clone detection. The accuracy of ChatGPT is average for the tasks of commit message generation, code review generation, natural language code search, merge conflict resolution. ChatGPT can be used for these tasks but perhaps the users will have to check its output carefully. ChatGPT performs poorly on code vulnerability detection. For certain tasks, like information retrieval based test prioritization, ChatGPT was not able to provide an answer at all.

Overall, ChatGPT does represent a very significant milestone and it can be used for some software engineering tasks \emph{as is} or for certain tasks as an adjunct tool.

\begin {comment}
# experiment results

Experiment status:

1.MethodName : done twice and transcript downloaded to chatgpt  exporter folder on one drive

2.code review : downloaded orignal set of experiments but it has only 7, so need to add couple more and download again

3. anaphora - downloaded and also more than 10 done.

4. commit msg - downloaded  and 10 done

5. log summary - downloaded transcript but i have asked only 6 logs to summarize.

#MethodName/Summarization

From code2seq website methods:
1. saveBitMapToFile - Correct by ChatGP/T

2. containsIgnoreCase - correct by chatgpt

3. addChildRequest - not quite correct, it said trackChildReq etc

4. get power of two - initially when i tried it had failed, saying it did not understand and gave literal translation, but this time, it was correct and said findNextPowerOfTwo 

5. generate prime number - chatgpt even better, suggested, generateRSAPrime (i did not get where RSA came from, till i saw the exception msg unable to gen prime for rsa key!, magnificent, even better than state of art)

6. compute stddev - chatgpt again slightly better in that it expanded stddev to standard deviation

7.index of target - chatgpt gave indexOf etc which is fine when you combine with the param name of target.

8. index of item - indexOfChild, findChildIndex etc, which is ok

9. check done - chatgpt better, it gave, waitUnTillFinished, blockTillDOne, which is better than "check done"

10. report entity size limit - chat gpt much much better, it said " 'checkEntitySizeLimits' - This name clearly conveys the purpose of the method, which is to check that the size of an entity falls within certain limits. " PERFECT!

11. wait for job - waitForJobCompletion - chatGPT better, the code2seq does NOT say whether we are wiating for job to begin or to end etc.

ALSO: chatgpt noticeably faster than the code2seq site!

-----Anaphora Ambguity-----
icse 22 anaphora paper
dataset xls DAMIR_Requirement_Anaphora_resolution

rs0-2 correct by chatgpt
rs0-3 correct
rs0-4 correct
rs0-8 correct
rs0-11 correct
rs0-19 corect
rs17-166b corect
rs16-53a corect
rs14-5 not sure, chatgpt said "it" refers to "onboard" but it also said "it" in second sentence refers to "etcs"
rs0-28 correct, the "it" refers to "ca and ta", the choices could have been "ca" or "ta" alone, but chatgpt did it correctly.

=====

type inference

it failed 2 of 9 times, because once return was a call to another function, and in another case, even the latest icse/fse 22 paper failed (node/abstract node).

1. parse : cgpt ok both param and return=>2/2
Ground Truth:
Arg: str
return:
Tuple[List[int, Placeholder],
Dict[str, Placeholder]]

cgpt ok - we have to count this as 2 correct of/2, one for input param and one for return.

2. _none_value cgpt ok 

this is from a discussion example in the allamanis typilus paper
ground truth is Optional[Union[float, int, str]].

cgpt didn't give the optional but it did say the
return was int, float, str etc depending on the if conditions in code. so cgpt ok

3. filter_wireless_controller_vap_data cgpt ok

this is from a discussion example in the allamanis typilus paper
ground truth is dict(str,Any)

4. _append_element cgpt notok - said it couldn't give type

from the icse22 paper discussion example

5. numpy_topk cgpt said its numpy array although its close mx.nd.NDArray, not sure it is ok/not ok

from the typilus paper

6. sample
not sure where this is from icse22paper ref 1 but its not there in the link?

7 get_transition
not sure where this is from 

8 createJobFromText
not sure where this is from  also unfair as return is a funciton call

9 un_pkcs_1_5 cgpt 2/3
from https://github.com/aldur/MatasanoCrypto/blob/master/matasano/blocks.py 

input params were int, int and return was bytes

10 forward cgpt ok and wonderfully too! 

in the above method, the return type is a dictionary. The key type is a string, and the value type is an instance of the torch.Tensor class.

from allamanis paper footnote
https://github.com/allenai/allennlp/pull/3376/files

cgpt said:
The input parameters of the function un_pkcs_1_5 are:
   b: a bytes-like object
   size: an integer
The return type of the function is bytes.

=====

code review automation:

 CODE REVIEW EXPERIMENTS DONE FROM FILE:
 
C:\Users\esrigir\Downloads\perfect_predictions\perfect_predictions\small\T5_pre-trained\code&comment-to-code\source_1.txt
 lines from above file which are used
 (each line has a java method and human code review technical_language tag has the actual review comment
comment & code in the above folder somewhere is review having both a comment and code on how to correct.
)

Lines:
1 chatgpt ok,
59 chatgpt not-ok, only after prompting it said log.debug should be log.error
3 cgpt ok, it did not say it but generated code returned the rule obj directly as desired by human review
10 cgpt not-ok, human review said 'exception not thrown' but cgpt did not say it
11 cgpt not-ok, human review said 'parts[1] is empty' but cgpt did not say it or reflect in code
43 cgpt ok - great behaviour!, human said the if(X) should be if(!X), cgpt did the fix as if (!X)
17 not sure whether chatgpt was right or wrong, human said use 'this.X=Y" but its not clear
22 cgpt not ok, human said "catch(X) {throw X}" doesn't make sense, only after prompting chatgpt said no point in catching X and rethrowing X.
42 cgpt ok, it got the human comment of saying no need for new Integer(id) autobox will take care of it.
63 cgpt nok, human review:"if (tasks.size()==0) return" unnecessary as next statement for loop won't anyway iterate. But cgpt said its ok, so not correct i think cgpt is.
86 cgpt nok, human review said use 'parseInt as return type was int instead of valueOf which gives Integer'

======

defect vulnerability detection:

data: C:\Users\esrigir\OneDrive - Ericsson (1)\chat-gpt-transcripts\codexglue-defect-detection-in-code.json

target 1 means vulnerability 0 means non-vulnerable

dataset from Devign paper
and
https://github.com/microsoft/CodeXGLUE/tree/main/Code-Code/Defect-detection

transcript copy pasted from chatgpt exporter at
C:\Users\esrigir\OneDrive - Ericsson (1)\chat-gpt-transcripts\chat-gpt-exporter\defect-vulnerability-devign.txt

r3d_read_rdvo actual 1 (vulnerable) chatgpt said 0 (wrong)
check_lowpass_line 1 0 wrong
vdadec_init 1 0 wrong
add 1 1 correct (overflow)
filter_mirror_setup 1 0 wrong
sub64 1 1 correct
assert_avoptions 1 1 correct
net_init_tap 1 0 wrong
ff_af_queue_init 0 1 wrong
emulated_push_error 1 1 correct
split_init 1 0 wrong

=====

Assert gen:

Data from old and new dataset at
C:\rs\chat-gpt-paper\experiments\assertion-generation-icse22

1. wrong (chatgpt, used variable in focal method in assert of test)
dev gold set: Assert . assertNotNull ( tasks )
2. right (
dev: assertFalse ( p1 . equals ( null ) )
3.ok
dev:assertTrue ( ttm . toString ( ) . contains ( ",010203.45," ) )
4.correct
dev:assertThat ( breeds , hasValue ( "gypsy" ) )
5.wrong?
dev:assertThat ( result . isValid ( ) , is ( false ) )
6.unfair-ignore
dev:Assert . assertEquals ( vector . dotProduct ( new Vector ( 20 , 25 , 30 ) ) , 800 , 1 )
7.correct
dev:assertThat ( isValid , is ( false ) )
8.correct
dev:assertEquals ( Arrays . asList ( "aa" , "ab" , "ac" ) , kittens )
9.correct (although chatgpt said canShowAds())
dev:assertTrue ( brutalAds . shouldShowAds ( ) )
10.wrong (chatgpt opposite):assertTrue(m.matches());
dev:assertEquals ( false , m . matches ( ) )

=====
clone detection

data in ericsson one drive folder under chat-gpt-transcripts with files of the form codex-clone-det-big-clone-bench-*"

Inputs given by me and the gold set answer (all pairs are clones according to gold set) are in the file under 
C:\Users\esrigir\OneDrive - Ericsson (1)\chat-gpt-transcripts\codex-clone-det-bigclone-my-chatgpt-inputs.txt

=====

Merge Conflict Resolution

Data from:C:\rs\chat-gpt-paper\experiments\fse-2022-MergeBERT-data.tar\fse-2022-MergeBERT-data\fse2022\automated-analysis-data\Java

files are of the form 20_a,20_b,20_base,20_merged,20_resoved
a,b are the two versions, base is the ancestor from which a,b diverged. merged has the merge conflicts, resolved has the developer resolved final code (ground truth or gold set).

Results:
20*files cgpt ok (copyrioght merge)
34* cgpt nok? i suppose they moved a class from one pkg to another and then modified the import, cgpt told to import one and use another with qualified name (impressive but not what the developer did in the resolved version)
54* cgpt nok, also imports but cgpt said it can't resolve merge conflicts as its a lang model
204698* cgpt ok, enum merge.
238341* bit confusing as base was empty and constansts were added to both versions
92165* confusing
resolved had "public enum DistType MULTINOMIAL, NORMAL, MULTINOMIAL_LOGISTIC, INDICATOR, INV_GAMMA; while cgpt correcty IMO included gaussian from base and version a. 
176737* cgpt ok, i think it did well, to bring latest version 3.5.1 as opposed to 3.5 in base from B and 2 new fields from A into final version
136052* cgpt nok, it did union of diff calls in A and B but really difficult as developer chose only one in A (requires domain)
142646* cgpt ok, matched dev resolved by adding an extends and new method 
131143* cgpt ok, but how did it do it?
87252* cgpt nok, method had diff return type so it said cant merge.
59908* cgpt seems ok but how did 

=====
refactor (extract method)

=====

\end{comment}

\bibliographystyle{ACM-Reference-Format}
\bibliography{refs}

\end{document}